**Title**
Probabilistic Social Learning Improves the Public's Detection of Misinformation

**Authors**
D. Guilbeault,[1]* S. Woolley,[2] J. Becker[3]

**Affiliations**
[1]Haas School of Business, University of California, Berkeley, USA 94720
[2]School of Journalism, University of Texas Austin, USA 78712
[3]School of Management, University of College London, UK E14 5AA

**\*Corresponding author:** Douglas Guilbeault (douglas.guilbeault@haas.berkeley.edu)

**Abstract**
The digital spread of misinformation is one of the leading threats to democracy, public health, and the global economy. Popular strategies for mitigating misinformation include crowdsourcing, machine learning, and media literacy programs that require social media users to classify news in binary terms as either true or false. However, research on peer influence suggests that framing decisions in binary terms can amplify judgment errors and limit social learning, whereas framing decisions in probabilistic terms can reliably improve judgments. In this preregistered experiment, we compare online peer networks that collaboratively evaluate the veracity of news by communicating either binary or probabilistic judgments. Exchanging probabilistic estimates of news veracity substantially improved individual and group judgments, with the effect of eliminating polarization in news evaluation. By contrast, exchanging binary classifications reduced social learning and entrenched polarization. The benefits of probabilistic social learning are robust to participants' education, gender, race, income, religion, and partisanship.

**Introduction**
The term *fake news* – defined as deliberately falsified news – has proliferated since the 2016 U.S. election (*1–7*). The practical risks associated with fake news have become increasingly apparent amid the COVID-19 pandemic and the upcoming 2020 U.S. election. A popular assumption of fake news research is that news can be effectively categorized in binary terms as either "real" or "fake" (*1–8*). Social media interventions often adopt this binary logic by using human crowdsourcing or machine learning to flag media as true or false, and by directing users to fact-checking websites that apply these binary classifications (*1–7*). Meanwhile, binary classifications of news veracity can fuel partisan conflict, as both right and left-wing media outlets regularly accuse the other of espousing fake news (*8*).

One challenge facing policymakers is that media literacy interventions involving binary classifications of news veracity report inconsistent effects (*1–7*). While some studies find that individuals can accurately classify news in binary terms (*1,4,5*), other studies suggest that individuals exhibit substantial biases in their news classifications (*2,3,8–13*), with the popular expectation that communication in online social networks leads to the rapid spread of inaccurate judgments (*7, 14–18*). Since social media users frequently discuss news in online peer networks (*6,7,19*), there is an urgent need to understand whether social influence exacerbates the spread of misinformation. Yet, prior work on misinformation relies primarily on observational data that is limited in isolating the causal effects of peer to peer communication on the public's capacity to evaluate news veracity (*14, 16–18*).

To the contrary, recent experimental work on collective intelligence suggests that communication in structured online networks can enable *social learning*, which occurs when exchanging information with peers improves belief accuracy (*20–23*). Importantly, this work has identified communication style as a key variable in determining whether social influence promotes social learning. Consistent with the theory that communication networks amplify the spread of misinformation, research on collective intelligence suggests that communicating about decisions in binary terms can propagate errors and limit *social learning* (*24–26*). However, more recent experimental work suggests that these limitations can be overcome when people communicate their beliefs in probabilistic terms. Experimental studies show that exchanging probabilistic judgments in online peer networks can improve belief accuracy in traditional estimation tasks (*20,26*), and also on more politicized topics, such as public health (*23*) and partisan policy (*21,22*). As a mechanism, it has been found that communicating beliefs using probabilistic estimates allows people to gauge uncertainty in their peers' judgments, which can incentivize belief change (*20–23*). Comparatively, exchanging binary classifications (such as whether someone identifies news as simply true or false) does not directly indicate individuals' confidence in their assessments. Furthermore, communicating beliefs using probabilistic estimates makes it easier for people to signal minor adjustments to their beliefs that can, at the population level, support significant improvements in belief accuracy (*20–23*).

In this preregistered experiment, we predict that allowing people to exchange probabilistic judgments in online peer networks will significantly improve their ability to accurately evaluate news veracity, as compared to peer networks that collectively evaluate news in binary terms, i.e. as true or false. (See *Supplementary Appendix* for details on pre-registration, https://osf.io/53b7v).

## Materials and Methods

900 subjects from Mechanical Turk (Mturk) participated in this experiment (Fig. 1). While the Mturk population can occasionally pose concerns regarding sample generalizability, methodological research has found that Mturk subjects provide high quality data for predicting online social media behavior, as compared to traditional survey methods (*27,28*). Additionally, the Mturk population has been widely used in both misinformation studies (*1,3–5*) and collective intelligence experiments (*20–23*). Moreover, because our research tests a scientific theory of information exchange as a function of carefully controlled communication environments, we do not expect our outcomes to be heavily influenced by demographic variation.

Subjects were randomized into one of two conditions. In the "binary" condition, subjects answered the question "Is the content of this message true?" (options: yes/no). In the "probabilistic" condition, subjects answered the question: "On a scale of 0 to 100, what is the likelihood that the content of this message is true?" A single trial in each condition consisted of 20 subjects tasked with evaluating the veracity of news before and after being able to see the beliefs of the other subjects in their trial.

Subjects in both conditions provided responses three times for each news item. In Round one, subjects gave an independent response without viewing the judgments of their peers. In Rounds two and three, subjects were shown a summary of their peer network's responses from the previous round (Fig. 1). Subjects in the binary condition were shown the percentage of their peer network that evaluated the content as true and false. Subjects in the probabilistic condition were shown their network's average estimate of the likelihood that the content is true.

Each peer network completed this process for four unique news items. The order of questions was constant in each block of four unique news items (see Fig. S1 for design). 12 news items were used covering a range of topics including vaccines, domestic politics, and terrorism (Fig. 1). The stimuli represented a range of formats, including social media posts and front-page headlines. Following recent work (*1,4*), we used the binary truth classifications of each news item provided by the professional fact-checking organization Snopes to determine its correct classification. Each trial evaluated two true and two false news items. Subjects received a monetary reward based on the accuracy of their final answer for each news item, consistent with accuracy incentives among social media users (*1,4*).

To detect social learning, we measured changes in subjects' reported beliefs along two dimensions: first, for each condition, we examined whether individuals and groups revised their veracity judgments in the correct direction with respect to Snopes' classification; and second, we examined whether individuals and groups improved in the categorical accuracy of their veracity classifications. In the binary condition, an individual is considered categorically accurate if the individual's binary response matches the binary classification (true/false) provided by Snopes. In the probabilistic condition, we determined the categorical accuracy of subjects' classifications by binarizing their numeric estimates and comparing these binarized judgments to Snopes' classifications: an estimate above 50% indicated that the subject believed the content to be true (i.e. more likely to be true), and an estimate below 50% indicated that the subject believed the content to be false (i.e. more likely to be false). In the binary condition, group-level accuracy refers to whether the majority decision (true/false) of a peer network matches Snopes' classification; and in the probabilistic condition, group-level accuracy refers to whether the average probability estimate of a peer network is above 50% for "true" stimuli or below 50% for "false" stimuli.

Finally, we measured whether peer networks shaped subjects' trust in online content, which has been identified as a key source of partisan differences (*1–6*). In the binary condition, subjects indicated trust by voting "yes" when asked whether a news item is true. In the probabilistic condition, individuals indicated trust in their estimate of the likelihood that a news item is true (above 50% indicated the belief that content is more likely to be true). Since each trial in each condition viewed two true and two false stimuli, the accurate rate at which subjects should trust content at baseline in our experiment is 50%.

Our subject pool was 43% Democratic, 23% Republican, 26% Independent, and 7.6% without identification. Subjects were randomized to condition regardless of partisanship, so each peer network contained a random mixture of political identifications. Importantly, we selected news items that represented a balanced range of partisan perspectives. Supplementary figures S2 – S13 provide each news item used in this experiment, along with crowdsourced ratings of each news item's partisan slant and extent of political bias. Data was collected between November 30[th] and December 12[th], 2018.

We conducted our experiment using Empirica.ly (*29*). Pre-registered power tests indicated the need for 15 trials in the binary condition and 30 trials in the probabilistic condition. We adopted the minimal number of trials needed in each condition to minimize the amount of exposure to misinformation, in accord with the request of Northwestern University's IRB, which approved this study.

**Results**

Before social interaction, there was no significant difference in the likelihood of peer networks providing the accurate judgment of news veracity in the binary and probabilistic condition ($p<0.43$, Wilcoxon Rank Sum), as expected by randomization. However, Fig. 2*A* shows that individuals from all partisan orientations were more likely to revise their judgments of news veracity in the correct direction when exchanging probabilistic rather than binary judgments ($p<0.001$, Wilcoxon Rank Sum). These results are robust to controlling for participants' gender, race, religiosity, income, strength of partisanship, and education, as well as the specific news items they evaluated ($p<0.001$, OR = 7.75, N=3190, SEs clustered by trial and condition). This finding similarly holds at the group-level (Fig. 2*B*, $p<0.001$, Wilcoxon Rank Sum).

Furthermore, the probabilistic condition was more effective at promoting improvements in the categorical accuracy of news classifications. Fig. 2*C* shows that the probabilistic condition significantly increased the likelihood that initially incorrect subjects would improve in the accuracy of their veracity classifications, compared to the binary condition ($p=0.05$, Wilcoxon Rank Sum); and Fig. 2*C* also shows that the probabilistic condition even more prominently increased the likelihood that initially incorrect groups would improve in the accuracy of their collective veracity classifications, compared to the binary condition ($p<0.01$, Wilcoxon Rank Sum). Consistent with the canonical wisdom of the crowd effect, Fig. 2*D* indicates that the benefits of probabilistic social learning are significantly more pronounced at the group-level than the individual-level ($p<0.05$, Wilcoxon Rank Sum).

Fig. 3 shows that the group-level benefits of probabilistic social learning were replicated across all topic areas represented by our stimuli (Figs. S2 – S13). Communication in the probabilistic condition led to significantly greater improvements in collective accuracy than the binary condition across five out of the six topic areas examined, including politics, economics, vaccines, terrorism, and domestic news ($p<0.05$, Wilcoxon Signed Rank Test). For the sixth topic area focusing on health, the probability of groups improving was identical in the binary and probabilistic condition. The benefits of probabilistic social learning were particularly pronounced for the topic areas of terrorism and news, where probabilistic communication led to over a 50 percentage-point increase in the likelihood of groups improving in categorical accuracy relative to the binary condition, which failed to enable any groups to improve for these topic areas (Fig. 3).

Critically, probabilistic communication not only increased belief accuracy, but also significantly reduced partisan differences (Fig. 4). Each group in each condition viewed two stimuli that were false and two stimuli that were true, such that at baseline, half of the content should have been evaluated as true and half should have been evaluated as false. Before peer interaction, Republicans were more likely to trust online content than Democrats in both the binary (Fig. 4*A*, median difference of 15 percentage points, $p<0.01$, Wilcoxon Rank Sum) and probabilistic (Fig. 4*B*, median difference of 7 percentage, $p<0.01$, Wilcoxon Rank Sum) condition. Supplementary analyses show that these baseline differences in trust reflect partisan biases in media evaluation, since these differences in trust correlate with the partisan slant of stimuli (Table S1, S2).

After communication in the binary condition, these partisan differences remained entrenched (Fig. 4C, $p<0.001$, Wilcoxon Rank Sum), and indeed nominally increased (by 0.7 percentage points). By the end of the task in the Binary condition, Republicans continued to be more likely to trust online content (Fig. 4C, median difference of 14 percentage points, $p<0.001$, Wilcoxon Rank Sum), leading to greater inaccuracy than Democrats (Fig. 4C, $p<0.01$, Wilcoxon Rank Sum). By comparison, communicating probabilistic judgments significantly reduced partisan biases in trust assessments (Fig. 4D, reduction of 5 percentage points, $p=0.01$, Wilcoxon Rank Sum), such that Republicans and Democrats no longer differed in their judgments of news veracity (Fig. 4D, $p<0.12$, Wilcoxon Rank Sum). As a result, in the probabilistic condition, both Democrats and Republicans converged on the accurate rate at which the stimuli should be trusted – that is, 50% of the time, since half of the stimuli were true and half were false for all groups.

**Discussion**

In popular applications of crowdsourcing (*18*) in misinformation detection (*1–7*), communication among human coders is frequently assumed to spread inaccurate judgments. Here we show that communication in online social networks systematically improves both individual and group judgments of new veracity. We observed these improvements both when subjects communicated using binary classifications as well as probabilistic judgments. Yet crucially, we show that the popular binary approach to classifying news as simply true or false can limit social learning when people communicate in online peer networks. We find that both individual and group-level judgments are more likely to improve when people communicate about news veracity using probabilistic judgments. Furthermore, we find that probabilistic communication can reduce partisan differences in news evaluation that otherwise remain entrenched when news veracity is socially evaluated in binary terms.

These results support the theory that exchanging probabilistic judgments is more effective at facilitating social learning than exchanging binary classifications. A limitation of our approach is that the underlying cognitive structure of individuals' beliefs remains a black box. It is possible that even in the binary condition, subjects' beliefs are internally represented in a probabilistic manner, such that their internal belief updating reflects similar dynamics to those observed in the probabilistic condition (*30,31*). However, if this is the case, this lends further support to our argument that social learning is improved by making the probabilistic nature of judgments explicit, because doing so allows subjects to more easily engage in dynamic belief updating that better reflects underlying belief structure, leading to greater improvements in the individual and collective classification of misinformation.

A central strength of our study is that we selected stimuli and procured social groups where the partisan identity of subjects was not explicitly salient, allowing us to experimentally isolate the effect of communication style on the capacity for social learning to improve misinformation classification. This is an appropriate set of experimental controls to impose, given that the identity of human coders is frequently anonymous in online crowdsourcing (*1,16,32,33*). Related work has shown that probabilistic communication is surprisingly robust at facilitating social learning across a range of partisan issues, in both politically-mixed (*21*) and politically-homogeneous social networks (*22*), and even in cases where the social identity of subjects is salient (*21–23*). These studies suggest that a promising direction for future research is to demonstrate the ability for probabilistic social learning to improve misinformation detection even in highly polarized political environments (*9,15,21,34*). Together, these results suggest that fact-checkers and social media organizations can better mitigate polarization and the spread of misinformation by using probabilistic representations of news veracity.

**Acknowledgments**

**General**:
D.G. and J. B. thank Damon Centola for extended discussions about social learning, and to Ned Smith for structural support in conducting this research. D.G. acknowledges support from a Joseph-Bombardier PhD scholarship from the Social Sciences and Humanities Research Council of Canada and a Dissertation Award from the Institute on Research and Innovation in Science. S.W. acknowledges support from the Knight Foundation and Omidyar Network as well as research support from the Center for Media Engagement (CME) and the School of Journalism at the University of Texas at Austin.

**Funding:**
The authors acknowledge financial support from the New Venture Fund.

**Author contributions:**

D.G., S.W., and J.B. designed the study. J.B. built the platform and collected the data. D.G. analyzed the data. D.G., S.W., and J.B. wrote the manuscript.

**Competing interests:**
There are no competing interests associated with this study.

**Data and materials availability:**
All data and code for statistical analyses are publicly available at
https://github.com/drguilbe/misinfoCI

**Figures and Tables**

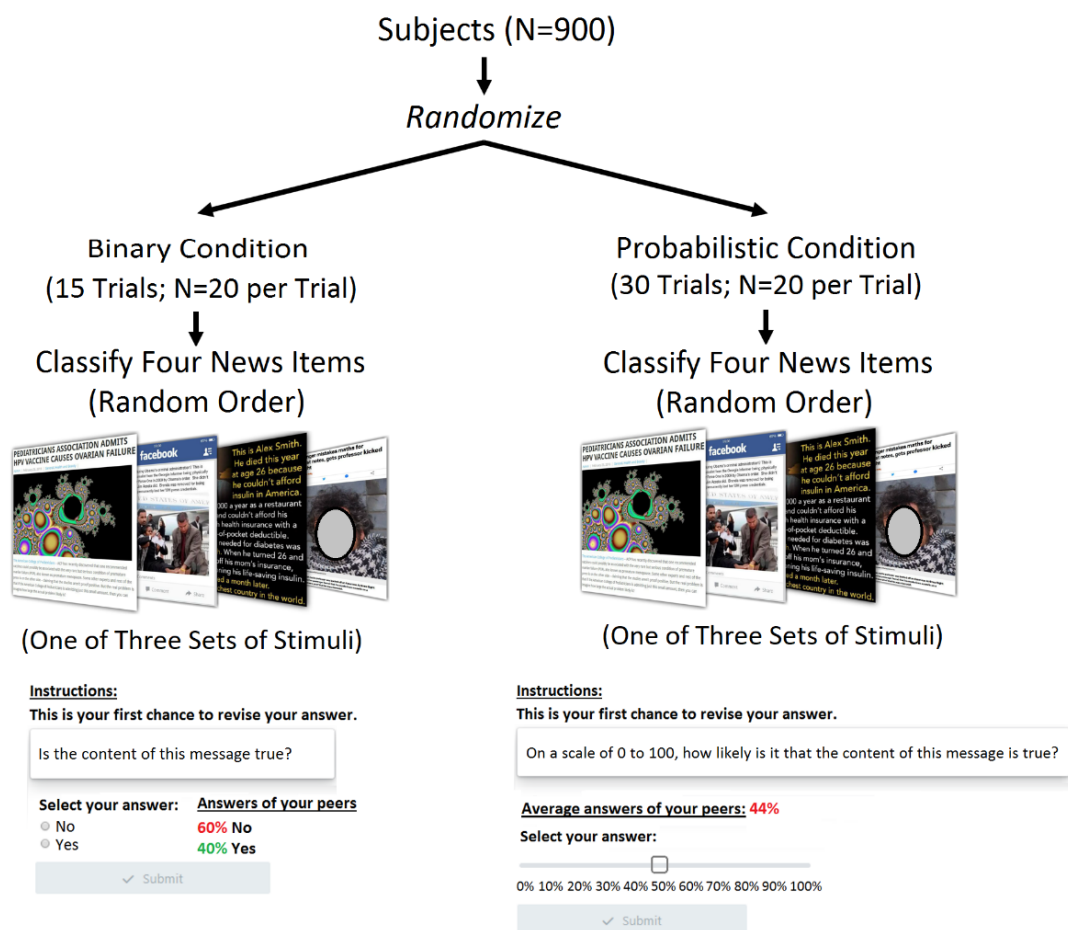

**Fig. 1. Experimental design.** Subjects were randomly assigned to a peer network that judged the veracity of news by exchanging either (i) binary or (ii) probabilistic judgments. Fewer trials were needed in the binary condition with equivalent statistical power ("Materials and Methods"). Each group in each condition consisted of 20 unique individuals. All group-level observations are independent.

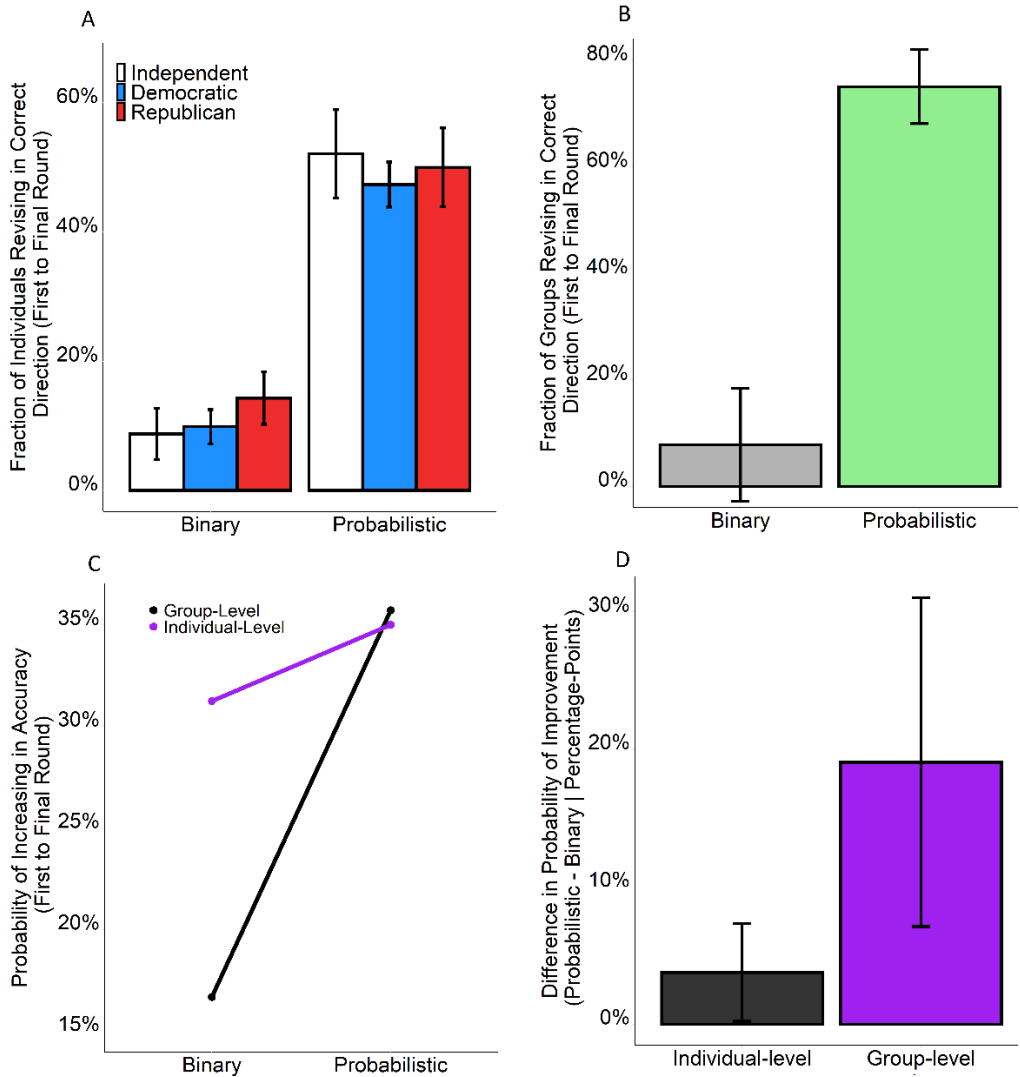

**Fig. 2. Comparing conditions in terms of the benefits of social learning for individual and collective judgments of news veracity.** (A) The fraction of subjects revising their judgments of news veracity in the correct direction, first to final round, split by partisanship and condition (averaged separately for each partisan group in each peer network). (B) The fraction of groups revising their collective judgments of news veracity in the correct direction, first to final round. (C) The probability of individuals and groups providing the correct veracity judgment by the final round, conditional on being initially incorrect at the first round. (D) The difference in the probability of individuals and groups improving in the accuracy of their veracity judgments between the binary and the probabilistic condition. (A, N=270; B, N=45; C, N=90; D, N=90). Error bars indicate 95% confidence intervals.

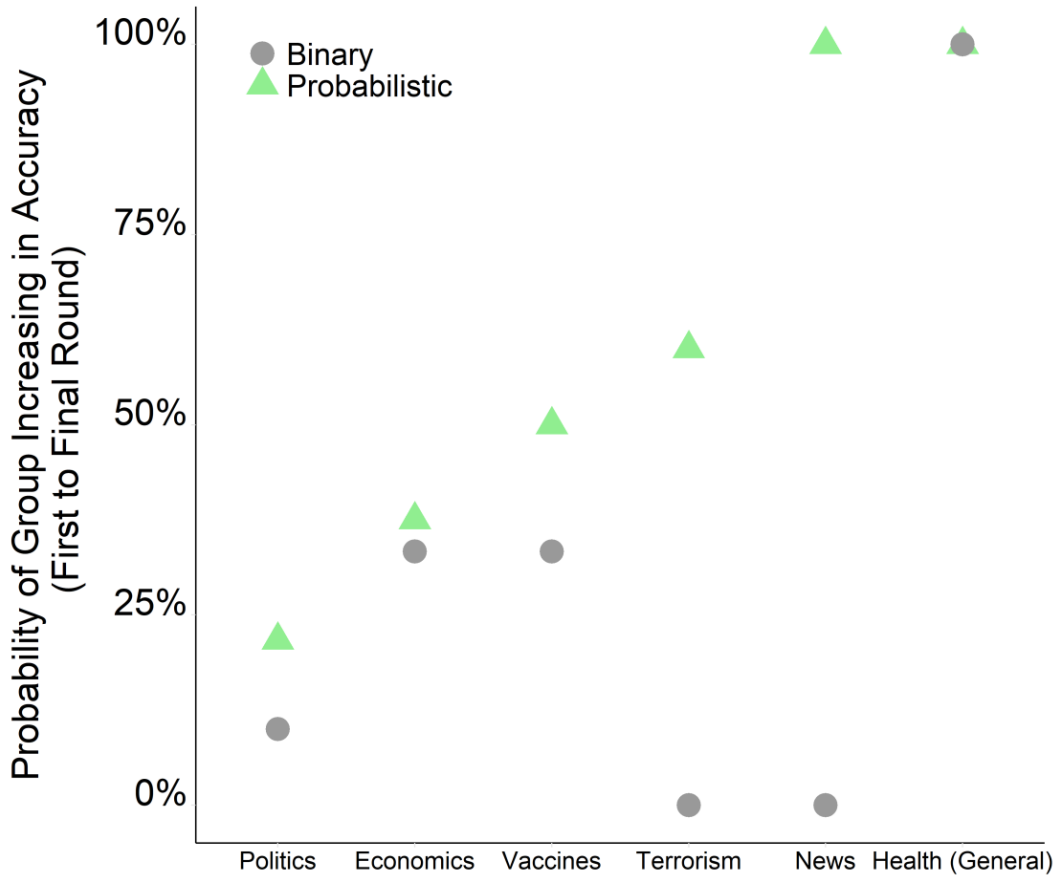

**Fig. 3. Comparing the binary and probabilistic condition in terms of the likelihood of groups increasing in the accuracy of their collective judgments, split by the topic area of news content.** The probability of groups increasing in accuracy is measured as the probability of providing the correct veracity judgment by the final round, conditional on the majority being initially incorrect at the first round. We calculate the fraction of groups that increased in accuracy for each question in each condition, and then we average this fraction by topic area. There were 9 questions in each condition with groups that were initially inaccurate, producing 18 question-level observations. The politics and vaccines topic each contained 3 questions; all other topic areas contained a single question.

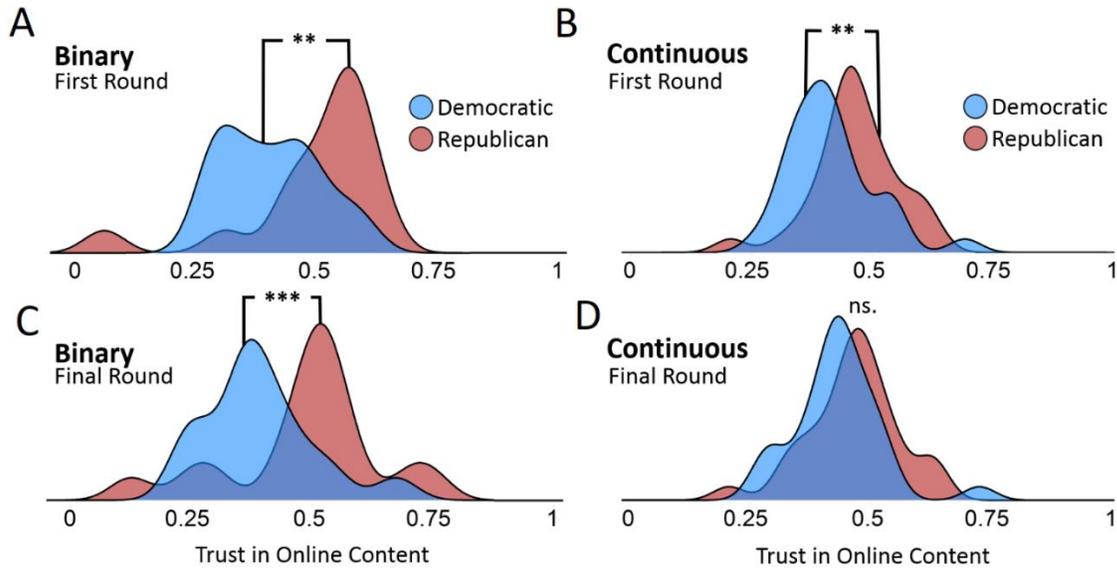

**Fig. 4. Partisan differences in trust toward online content, averaged across questions at the trial-level.** Trust is measured by the rate at which subjects in each condition evaluated questions as more likely to be true. Density distributions indicate the fraction of questions that each trial collectively evaluated as true according to the communication style of each condition (binary vs. probabilistic). The rate at which questions were evaluated as true was averaged separately for both partisan groups in each trial. The data display the fraction of questions that Democrats and Republicans in each trial evaluated as true for the binary condition at the first (Panel A) and final round (Panel C), and for the probabilistic condition at the first (Panel B) and the final round (Panel D). Since each group in each condition encountered two true and two false news items, the appropriate fraction of questions that each trial should evaluate as true is 50%. Panel A & C, N=30; C & D, N=60. *$p<0.1$; **$p<0.01$; ***$p<0.001$; ns., not significant.